\documentstyle[12pt]{article}
\begin{document}

\def\PRL#1{{\it Phys.~Rev.~Lett.~}{\bf#1}}
\def\PRD#1{{\it Phys.~Rev.~}{\bf D#1}}
\def\NPB#1{{\it Nucl.~Phys.~}{\bf B#1}}
\def\PLB#1{{\it Phys.~Lett.~}{\bf B#1}}

\def\Slash{\hskip -.6em/}
\def\ttbar{t\bar{t}}
\def\bbbar{b\bar{b}}
\def\btbar{b\bar{t}}
\def\tbbar{t\bar{b}}
\def\ImPi{{\rm Im}\Pi}
\def\Imlambda{{\rm Im}\lambda}
\def\ImDelta{{\rm Im}\Delta}
\def\Imf{{\rm Im}f}
\def\MSbar{\overline{\rm MS}}

\def\beq{\begin{equation}}
\def\eeq{\end{equation}}
\def\beqa{\begin{eqnarray}}
\def\eeqa{\end{eqnarray}}

\def\cf{{\it cf.}}
\def\ie{{\it i.e.}}

\begin{flushright}
FERMILAB-PUB-94/172-T \\
EFI 94-32 \\
July 1994 \\
\end{flushright}

\bigskip
\medskip

\begin{center}

\large

{\bf On the Choice of Dispersion Relation to Calculate
the QCD Correction to $\Gamma(H\rightarrow \ell^+\ell^-)$.}

\bigskip
\medskip

\normalsize

\bigskip
{\it Tatsu Takeuchi \\
\medskip
Fermi National Accelerator Laboratory \\
P.O. Box 500, Batavia, IL 60510 } \\
\bigskip

and

\bigskip
{\it Aaron K. Grant and Mihir P. Worah} \\
\medskip
{\it Enrico Fermi Institute and Department of Physics \\
University of Chicago \\
5640 S. Ellis Avenue, Chicago, IL 60637} \\
\bigskip

\bigskip


{\bf ABSTRACT} \\
\end{center}

\begin{quote}

We use the Operator Product Expansion (OPE) of quark vacuum
polarization functions to show that the dispersion relation of
Kniehl and Sirlin
will yield the correct result to all orders in $\alpha_s$ when
applied to the QCD correction to the leptonic decay width of the
Higgs boson.

\end{quote}

\newpage

\centerline{\bf I.  INTRODUCTION}
\bigskip

Dispersion relations (DR's) are widely used to calculate higher order
electroweak radiative corrections \cite{Chang,DRcalc}.
However, the freedom to make subtractions in DR's often makes the
choice of one DR over another a delicate issue
\cite{KSDR,Madison},
since a certain DR may work for some observables, but not for others.
The use of a particular DR is often justified by comparing the result
with that of a direct calculation using dimensional regularization
or some other computational technique at a given order \cite{KScomp}.

We wish to emphasize that in so far as higher order QCD
corrections are involved, the Operator Product Expansion (OPE)
can be used to decide on a DR without explicitly
doing a multi--loop calculation.
In a previous paper \cite{TGW} we illustrated this with
the example of higher order corrections to $\Delta\rho$.  There we found
that the subtracted DR's employed in Refs.~\cite{Chang,KSDR}, as well as
the na\"\i ve unsubtracted DR give the correct result for $\Delta\rho$.
In this paper, we wish to illustrate this use of the OPE
by considering corrections to the decay width
$\Gamma(H\rightarrow\ell^+\ell^-)$, and show that in this particular
case one does need a subtracted DR, and that furthermore
the subtraction of Ref.~\cite{KSDR} gives the correct result, whereas
that of Ref.~\cite{Chang} does not.

This paper is organized as follows.  In Section II, we discuss the
role of subtractions in DR's and the use of the OPE in determining
the correctness of the subtraction. In Section III we show that the
DR of Ref.~\cite{KSDR} gives the correct answer to all orders in $\alpha_s$
when used to calculate the QCD correction to
$\Gamma(H\rightarrow \ell^+\ell^-)$.
In Section IV we show that the DR
of Ref.~\cite{Chang} gives the incorrect answer for the same correction.
Section V concludes.

\bigskip
\bigskip

\bigskip
\centerline{\bf II. THE NECESSITY AND CORRECTNESS OF SUBTRACTIONS}
\bigskip

In this section, we will look at the self energies that
must be computed to obtain the QCD correction to
$\Gamma(H\rightarrow\ell^+\ell^-)$.   We will show that a na\"\i ve
application of Cauchy's theorem will not lead to a
dispersion relation for this correction and that a subtraction
must be introduced.   We then discuss how one may check which
choice of subtraction is the correct one.

QCD corrections to the decay width
$\Gamma(H\rightarrow \ell^+\ell^-)$ enter through the
quark contribution to the self energies of the $W$ and the Higgs.
In Ref.~\cite{Kniehl}, it was shown that
\beq
\Gamma(H\rightarrow \ell^+\ell^-)
= \Gamma_0(H\rightarrow \ell^+\ell^-)
  \Bigl[ 1 + \delta + \cdots \Bigr],
\eeq
where
\beq
\delta = \frac{\Pi_{WW}(0)}{M_W^2} + \Pi'_{HH}(M_H^2).
\label{deldef}
\eeq
The tree level value $\Gamma_0$ is assumed to be expressed in
terms of $G_\mu$ which is why $\Pi_{WW}(0)$, the self--energy of the $W$
evaluated at zero momentum transfer,
appears in this formula.  The derivative of
the self--energy of the Higgs, $\Pi'_{HH}(M_H^2)$, comes
from the wave--function renormalization constant of the
Higgs field.

In contrast to the $\Delta\rho$ case that was considered in a
previous paper \cite{TGW}, one must introduce a subtraction
in order to write down a dispersion relation for $\delta$.
This can be seen as
follows:
applying Cauchy's theorem to $\Pi_{WW}(s)$ and $\Pi_{HH}(s)$,
we can write
\beqa
\Pi_{WW}(s) & = & \frac{1}{\pi}\int^{\Lambda^2} ds'
                  \frac{ \ImPi_{WW}(s') }{ s'-s-i\epsilon }
              +   \frac{1}{2\pi i}\oint_{|s|=\Lambda^2} ds'
                  \frac{ \Pi_{WW}(s') }{ s'-s },
\nonumber    \\
\Pi_{HH}(s) & = & \frac{1}{\pi}\int^{\Lambda^2} ds'
                  \frac{ \ImPi_{HH}(s') }{ s'-s-i\epsilon }
              +   \frac{1}{2\pi i}\oint_{|s|=\Lambda^2 } ds'
                  \frac{ \Pi_{HH}(s') }{ s'-s }.
\label{Cauchy}
\eeqa
We consider the $\Pi(s)$'s to be regularized and
finite so that both sides of these equations are well defined.
Note that from this point of view, the radius of the contour
$\Lambda^2$ has
no relation to the ultraviolet regulator which makes the $\Pi(s)$'s
finite.
Substituting these expressions into Eq.~(\ref{deldef}), we find
\beq
\delta = \delta_0(\Lambda^2) + R_0(\Lambda^2),
\eeq
where
\beqa
\delta_0(\Lambda^2)
& = & \frac{1}{\pi}\int^{\Lambda^2} ds
      \left[ \frac{1}{M_W^2} \frac{ \ImPi_{WW}(s) }{ s }
           + \frac{ \ImPi_{HH}(s) }{ (s-M_H^2-i\epsilon)^2 }
      \right],
\nonumber \\
R_0(\Lambda^2)
& = & \frac{1}{2\pi i}\oint_{|s|=\Lambda^2} ds
      \left[ \frac{1}{M_W^2} \frac{ \Pi_{WW}(s) }{ s }
           + \frac{ \Pi_{HH}(s) }{ (s-M_H^2)^2 }
      \right].
\label{DR1}
\eeqa
Since $\Pi_{WW}(s)\sim s$ and $\Pi_{HH}(s)\sim s$ as $s\rightarrow\infty$,
both $\delta_0(\Lambda^2)$ and $R_0(\Lambda^2)$
diverge quadratically as
$\Lambda^2\rightarrow\infty$ even though their sum is finite and
independent of $\Lambda^2$.
Therefore, we find that $\delta$ cannot be replaced with $\delta_0(\infty)$,
and that the na\"\i ve substitutions of Eq.~(\ref{Cauchy})
do not lead to a dispersion relation which expresses $\delta$ as
an integral involving only the imaginary parts of the $\Pi(s)$'s.

This problem can be solved by noticing that the representation of
the $\Pi(s)$'s as an integral along the real $s$ axis plus an
integral around the circle at $|s|=\Lambda^2$ is not unique.
It is always possible to introduce an analytic function $f(s)$ such that
\beq
0 = \frac{1}{\pi}\int^{\Lambda^2} ds \;\Imf(s)
  + \frac{1}{2\pi i}\oint_{|s|=\Lambda^2} ds \;f(s),
\eeq
and write
\beqa
\Pi_{WW}(s)
& = & \frac{1}{\pi}\int^{\Lambda^2} ds'
      \left[ \frac{ \ImPi_{WW}(s') }{ s'-s-i\epsilon }
           + \Imf(s')
      \right]
\nonumber \\
& & + \frac{1}{2\pi i}\oint_{|s|=\Lambda^2} ds'
      \left[ \frac{ \Pi_{WW}(s') }{ s'-s }
           + f(s')
      \right],
\label{WWsub}
\eeqa
without changing the value of $\Pi_{WW}(s)$.
(We do not consider a similar subtraction on $\Pi_{HH}(s)$
here because the subtraction will not contribute to the
derivative $\Pi'_{HH}(s)$.)

Substituting Eq.~(\ref{WWsub}) into Eq.~(\ref{deldef})
gives us
\beq
\delta = \delta_f(\Lambda^2) + R_f(\Lambda^2),
\eeq
where
\beqa
\delta_f(\Lambda^2)
& = & \frac{1}{\pi}\int^{\Lambda^2} ds
             \left[ \frac{1}{M_W^2}
                    \left\{ \frac{ \ImPi_{WW}(s) }{ s }
                          + \Imf(s)
                    \right\}
                  + \frac{ \ImPi_{HH}(s) }{ (s-M_H^2-i\epsilon)^2 }
             \right],
\nonumber \\
R_f(\Lambda^2)
& = &  + \frac{1}{2\pi i}\oint_{|s|=\Lambda^2} ds
         \left[ \frac{1}{M_W^2}
                \left\{ \frac{ \Pi_{WW}(s) }{ s }
                       + f(s)
                \right\}
              + \frac{ \Pi_{HH}(s) }{ (s-M_H^2)^2 }
         \right].
\label{DRsub}
\eeqa
If the subtraction $f(s)$ is judiciously chosen so that
\beq
\lim_{\Lambda^2\rightarrow\infty} R_f(\Lambda^2) = 0,
\label{Condition}
\eeq
then one obtains the dispersion relation
\beq
\delta = \delta_f(\infty),
\label{DRf}
\eeq
{\it i.e.} we can represent $\delta$ by an integral over the real $s$
axis only.

We would like to emphasize here the viewpoint that the
purpose of subtractions
in any DR is to make the integral around the circle at $|s|=\Lambda^2$
disappear.   If the integral disappears automatically in the
linear combination
of vacuum polarization functions that one wishes to calculate, then
subtractions are not necessary.  For instance,
this has been shown to be the case for $\Delta\rho$ \cite{TGW}.

Whether a certain choice of the subtraction $f(s)$ is correct or
not can be checked in two ways.  The first is to
calulate $\delta_f(\infty)$
and see if it reproduces the correct result to a given order in
$\alpha_s$.
This was the strategy used in Ref.~\cite{KScomp}.  This technique is
useful for motivating the choice of one dispersion relation over another,
but cannot rigorously establish the correctness of such a choice to all
orders in $\alpha_s$.

A second  method is to see if the
condition ~(\ref{Condition}) is satisfied.
Since we only need to know the behavior of the
integrand in the limit $s\rightarrow\infty$, the operator
product expansion (OPE) will suffice to tell us whether this condition is
satisfied to {\it all} orders in $\alpha_s$.
This is the approach we will use in the following.

\newpage
\centerline{\bf III.  THE KNIEHL--SIRLIN SUBTRACTION}
\bigskip

In this section, we will look at the subtraction
introduced  in Ref.~\cite{KSDR} and show that
Eq.~(\ref{Condition}) is indeed satisfied.

Following Ref.~\cite{KSDR}, we define the following notation:
\beqa
\Pi^{V,A}_{\mu\nu}(q,m_1,m_2)
& = & -i \int d^4x e^{iq\cdot x}
\langle 0 | T^* \left[ J_\mu^{V,A}(x) J_\nu^{V,A\dagger}(0)
                \right]
| 0 \rangle                                            \nonumber \\
& = & g_{\mu\nu} \Pi^{V,A}(s,m_1,m_2)
    + q_\mu q_\nu \lambda^{V,A}(s,m_1,m_2)
      \phantom{\frac{1}{2}}                            \nonumber \\
& = & \left( g_{\mu\nu} - \frac{q^\mu q^\nu}{q^2}
      \right) \Pi^{V,A}(s,m_1,m_2)
    + \left( \frac{q^\mu q^\nu}{q^2}
      \right) \Delta^{V,A}(s,m_1,m_2),  \nonumber \\
\label{Defs1}
\eeqa
where $s = q^2$, and $J_\mu^{V,A}(x)$ represents
the vector and axial vector currents
constructed from quark fields, respectively.
Note that
\beqa
\Pi^{V,A}(s) = \Delta^{V,A}(s) - s\lambda^{V,A}(s),
\label{Defdelm}
\eeqa
so that
\beq
\Pi^{V,A}(0) = \Delta^{V,A}(0),
\eeq
unless $\lambda^{V,A}(s)$ has a pole at $s=0$.
We further introduce the notation
\beqa
\Pi^{V,A}_\pm(s)
& = & \Pi^{V,A}(s,m_1,m_2),     \phantom{\frac{1}{2}} \nonumber \\
\lambda^{V,A}_\pm(s)
& = & \lambda^{V,A}(s,m_1,m_2), \phantom{\frac{1}{2}} \nonumber \\
\Delta^{V,A}_\pm(s)
& = & \Delta^{V,A}(s,m_1,m_2),  \phantom{\frac{1}{2}} \nonumber \\
\Pi^{V,A}_0(s)
& = & \frac{1}{2} \left[ \Pi^{V,A}(s,m_1,m_1) + \Pi^{V,A}(s,m_2,m_2)
                  \right],       \nonumber \\
\lambda^{V,A}_0(s)
& = & \frac{1}{2} \left[ \lambda^{V,A}(s,m_1,m_1) + \lambda^{V,A}(s,m_2,m_2)
                  \right],       \nonumber \\
\Delta^{V,A}_0(s)
& = & \frac{1}{2} \left[ \Delta^{V,A}(s,m_1,m_1) + \Delta^{V,A}(s,m_2,m_2)
                  \right].
\label{Defs2}
\eeqa
The conservation of the neutral vector currents implies
the Ward Identities:
\beq
\Pi^V_0(s) = -s\lambda^V_0(s), \qquad\qquad \Delta^V_0(s) \equiv 0.
\label{WI}
\eeq

These definitions let us write
the contribution of a quark doublet, with masses $m_1$, and $m_2$,
to $\Pi_{WW}(0)$ as
\beq
\Pi_{WW}(0)  =  \frac{g^2}{8}
                \left[ \Pi^V_\pm(0) + \Pi^A_\pm(0)
                \right]
             =  \frac{g^2}{8}
                \left[ \Delta^V_\pm(0) + \Delta^A_\pm(0)
                \right].
\label{PiWW}
\eeq

The subtraction scheme of Ref.~\cite{KSDR}
is then given by,
\beqa
\Pi^{V,A}_{\pm,0}(s)
& = & \frac{1}{\pi}\int^{\Lambda^2} ds'
      \left[ \frac{ \ImPi^{V,A}_{\pm,0}(s') }{ s'-s-i\epsilon }
           + \Imlambda^{V,A}_{\pm,0}(s')
      \right]
\nonumber \\
&   &
     + \frac{1}{2\pi i}\oint_{|s|=\Lambda^2} ds'
        \left[ \frac{ \Pi^{V,A}_{\pm,0}(s') }{ s'-s }
           + \lambda^{V,A}_{\pm,0}(s')
        \right].
\label{KSsub1}
\eeqa
Using Eq.~(\ref{Defdelm}), this can also be written as
\beqa
\Pi^{V,A}_{\pm,0}(s)
& = & \frac{1}{\pi}\int^{\Lambda^2} ds'
      \left[ \frac{ \ImDelta^{V,A}_{\pm,0}(s')  }{ s'-s-i\epsilon }
         - s \frac{ \Imlambda^{V,A}_{\pm,0}(s') }{ s'-s-i\epsilon }
      \right]
\nonumber \\
&   &
      + \frac{1}{2\pi i}\oint_{|s|=\Lambda^2} ds'
        \left[ \frac{ \Delta^{V,A}_{\pm,0}(s')  }{ s'-s }
           - s \frac{ \lambda^{V,A}_{\pm,0}(s') }{ s'-s }
        \right],
\label{KSsub2}
\eeqa
which shows that the effect of the
subtraction amounts to applying Cauchy's theorem to
the $\Delta(s)$'s and $\lambda(s)$'s instead of
the $\Pi(s)$'s.

Application of this subtraction to $\Pi_{WW}(0)$ gives us
\beqa
\frac{\Pi_{WW}(0)}{M_W^2}
& = & \frac{G_\mu}{\sqrt{2}}
      \left[ \frac{1}{\pi}\int^{\Lambda^2} \frac{ds}{s}
             \left\{ \ImDelta^V_\pm(s) + \ImDelta^A_\pm(s)
             \right\}
      \right.
\nonumber \\
&   & \left.
           + \frac{1}{2\pi i}\oint_{|s|=\Lambda^2} \frac{ds}{s}
             \left\{ \Delta^V_\pm(s) + \Delta^A_\pm(s)
             \right\}
      \right].
\eeqa
Therefore, we can write
\beq
\delta = \delta_{KS}(\Lambda^2) + R_{KS}(\Lambda^2),
\eeq
where
\beqa
\delta_{KS}(\Lambda^2)
& = & \frac{1}{\pi}\int^{\Lambda^2} ds
      \left[ \frac{G_\mu}{\sqrt{2}}
             \left\{ \frac{ \ImDelta^V_\pm(s) }{ s }
                   + \frac{ \ImDelta^A_\pm(s) }{ s }
             \right\}
           + \frac{ \ImPi_{HH}(s) }{ (s-M_H^2-i\epsilon)^2 }
      \right],
\nonumber \\
R_{KS}(\Lambda^2)
& = & \frac{1}{2\pi i}\oint_{|s|=\Lambda^2} ds
      \left[ \frac{G_\mu}{\sqrt{2}}
             \left\{ \frac{ \Delta^V_\pm(s) }{ s }
                   + \frac{ \Delta^A_\pm(s) }{ s }
             \right\}
           + \frac{ \Pi_{HH}(s) }{ (s-M_H^2)^2 }
      \right].
\eeqa

In order to show that
\beq
\lim_{\Lambda^2\rightarrow\infty} R_{KS}(\Lambda^2) = 0,
\eeq
we need the following two relations.
The first is that when $s \gg v^2$,
\beq
\Pi_{HH}(s) = \Pi_{\chi\chi}(s)
              \left[ 1 + O\left( \frac{v^2}{s}
                          \right)
              \right],
\label{HandG}
\eeq
where $\Pi_{\chi\chi}(s)$ is the self-energy of the neutral Goldstone
boson (which is absorbed into the $Z$), and $v$ is the Higgs VEV.
This can easily be seem to be true since the two functions must
coincide in the limit $v^2\rightarrow 0$.
The second is the Ward Identity
\beq
\Pi_{\chi\chi}(s) = -\frac{G_\mu}{\sqrt{2}}s\Delta^A_0(s),
\label{WIiso}
\eeq
which comes from the conservation of isospin currents.
See Ref.~\cite{Barbieri}.

Using Eqs.~(\ref{HandG}) and (\ref{WIiso}), we find
\beqa
\lim_{\Lambda^2\rightarrow\infty}R_{KS}(\Lambda^2)
& = & \lim_{\Lambda^2\rightarrow\infty}
      \frac{1}{2\pi i}\oint_{|s|=\Lambda^2} ds
      \left[ \frac{G_\mu}{\sqrt{2}}
             \left\{ \frac{ \Delta^V_\pm(s) }{ s }
                   + \frac{ \Delta^A_\pm(s) }{ s }
             \right\}
           + \frac{ \Pi_{HH}(s) }{ (s-M_H^2)^2 }
      \right]
\nonumber \\
& = & \lim_{\Lambda^2\rightarrow\infty}
      \frac{G_\mu}{\sqrt{2}}
      \frac{1}{2\pi i}\oint_{|s|=\Lambda^2} \frac{ds}{s}
      \left[ \Delta^V_\pm(s)
           + \Delta^A_\pm(s)
           - \Delta^A_0(s)
      \right].
\eeqa

The asympotitic forms of the $\Delta(s)$'s
as $|s|\rightarrow\infty$ can be gleaned from their
OPE's found in the appendix of Ref.~\cite{Braaten}.
They are:
\beqa
\Delta^V_\pm(-Q^2)
& = & \hat{C}_{\Delta 1}(Q)
      \left[ \hat{m}_1(Q) - \hat{m}_2(Q)
      \right]^2
    + \hat{C}_{\Delta 2}(\mu)
      \left[ \hat{m}_1(\mu) - \hat{m}_2(\mu)
      \right]^2
    + O\left( \frac{1}{Q^2}
       \right),
\nonumber \\
\Delta^A_\pm(-Q^2)
& = & \hat{C}_{\Delta 1}(Q)
      \left[ \hat{m}_1(Q) + \hat{m}_2(Q)
      \right]^2
    + \hat{C}_{\Delta 2}(\mu)
      \left[ \hat{m}_1(\mu) + \hat{m}_2(\mu)
      \right]^2
    + O\left( \frac{1}{Q^2}
       \right),
\nonumber \\
\Delta^A_0(-Q^2)
& = & \hat{C}_{\Delta 1}(Q)
      \left[ 2\hat{m}_1(Q)^2   + 2\hat{m}_2(Q)^2
      \right]
\nonumber \\
& & + \hat{C}_{\Delta 2}(\mu)
      \left[ 2\hat{m}_1(\mu)^2 + 2\hat{m}_2(\mu)^2
      \right]
    + O\left( \frac{1}{Q^2}
       \right).
\label{DeltaOPE}
\eeqa
Though the OPE's are derived in the deep Euclidean region
$-s = Q^2 \gg 0$, the power dependence of the $\Delta(s)$'s on $s$
will be the same all around the circle at $|s| = \Lambda^2$.
Therefore, we can see immediately that
\beqa
\lim_{\Lambda^2\rightarrow\infty}R_{KS}(\Lambda^2)
& = & \lim_{\Lambda^2\rightarrow\infty}
      \frac{G_\mu}{\sqrt{2}}
      \frac{1}{2\pi i}\oint_{|s|=\Lambda^2} \frac{ds}{s}
      \left[ \Delta^V_\pm(s)
           + \Delta^A_\pm(s)
           - \Delta^A_0(s)
      \right]
\nonumber \\
& = & 0.
\eeqa
Therefore,
\beq
\delta = \delta_{KS}(\infty).
\label{KSDRx}
\eeq

We believe this derivation clarifies the reason why the
DR of Ref.~\cite{KSDR} was found to give the correct answer
at $O(\alpha\alpha_s)$ in Ref.~\cite{KScomp}.  In fact,
since the OPE is correct to all orders in $\alpha_s$,
Eq.~(\ref{KSDRx}) is also correct to all orders in $\alpha_s$.

\newpage
\centerline{\bf IV. THE SUBTRACTION OF CHANG, GAEMERS}
\centerline{\bf AND VAN~NEERVEN}
\bigskip

Next, we will look at the subtraction introduced in Ref.~\cite{Chang}
and show that in contrast to the
subtraction of Ref.~\cite{KSDR},
Eq.~(\ref{Condition}) is not satisfied.

The subtraction introduced in Ref.~\cite{Chang} is
given by
\beqa
\Pi^{V,A}_{0,\pm}(s)
& = &
  \frac{1}{\pi}\int^{\Lambda^2}ds'
  \left[ \frac{ \ImPi^{V,A}_{0,\pm}(s') }{ s'-s-i\epsilon }
       + \Imlambda^V_0(s')
  \right]   \nonumber \\
& &
+ \frac{1}{2\pi i}\oint_{|s|=\Lambda^2}ds'
  \left[ \frac{ \Pi^{V,A}_{0,\pm}(s') }{ s'-s }
       + \lambda^V_0(s')
  \right].
\eeqa
Note that the difference from the scheme of Ref.~\cite{KSDR} is that
the same subtraction $\lambda^V_0$ is used for all four cases,
$\Pi^V_0$, $\Pi^A_0$, $\Pi^V_\pm$, and $\Pi^A_\pm$.

This time, $\Pi_{WW}(0)$ is written as
\beqa
\frac{\Pi_{WW}(0)}{M_W^2}
& = & \frac{G_\mu}{\sqrt{2}}
      \left[ \frac{1}{\pi}\int^{\Lambda^2} ds
             \left\{ \frac{ \ImPi^V_\pm(s) }{ s }
                   + \frac{ \ImPi^A_\pm(s) }{ s }
                   + 2\Imlambda^V_0(s)
             \right\}
      \right.
\nonumber \\
&   & \left.
           + \frac{1}{2\pi i}\oint_{|s|=\Lambda^2} ds
             \left\{ \frac{ \Pi^V_\pm(s) }{ s }
                   + \frac{ \Pi^A_\pm(s) }{ s }
                   + 2\lambda^V_0(s)
             \right\}
      \right]
\nonumber \\
& = & \frac{G_\mu}{\sqrt{2}}
      \left[ \frac{1}{\pi}\int^{\Lambda^2} ds
             \left\{ \frac{ \ImDelta^V_\pm(s) }{ s }
                   + \frac{ \ImDelta^A_\pm(s) }{ s }
                   + 2\Imlambda^V_0(s)
                   - \Imlambda^V_\pm(s)
                   - \Imlambda^A_\pm(s)
             \right\}
      \right.
\nonumber \\
&   & \left.
           + \frac{1}{2\pi i}\oint_{|s|=\Lambda^2} ds
             \left\{ \frac{ \Delta^V_\pm(s) }{ s }
                   + \frac{ \Delta^A_\pm(s) }{ s }
                   + 2\lambda^V_0(s)
                   - \lambda^V_\pm(s)
                   - \lambda^A_\pm(s)
             \right\}
      \right].
\eeqa
This gives us
\beq
\delta = \delta_{CGN}(\Lambda^2) + R_{CGN}(\Lambda^2),
\eeq
where
\beqa
\delta_{CGN}(\Lambda^2)
& = & \delta_{KS}(\Lambda^2)
    + \frac{G_\mu}{\sqrt{2}}
      \frac{1}{\pi}\int^{\Lambda^2} ds
      \left[ 2\Imlambda^V_0(s)
            - \Imlambda^V_\pm(s)
            - \Imlambda^A_\pm(s)
      \right],
\nonumber \\
R_{CGN}(\Lambda^2)
& = & R_{KS}(\Lambda^2)
    + \frac{G_\mu}{\sqrt{2}}
      \frac{1}{2\pi i}\oint_{|s|=\Lambda^2} ds
      \left[ 2\lambda^V_0(s)
            - \lambda^V_\pm(s) - \lambda^A_\pm(s)
      \right].
\eeqa
Using the OPE's of the $\lambda(s)$'s, again from Ref.~\cite{Braaten}:
\beqa
\lambda^V_\pm(-Q^2)
& = & \hat{C}_{\lambda1}(Q)
    + \hat{C}_{\lambda2}(Q)\frac{[\hat{m}_1(Q) + \hat{m}_2(Q)]^2}{Q^2}
      \nonumber \\
&   & \qquad\qquad
    + \hat{C}_{\lambda3}(Q)\frac{[\hat{m}_1(Q) - \hat{m}_2(Q)]^2}{Q^2}
    + O\left( \frac{1}{Q^4} \right),
      \nonumber \\
\lambda^A_\pm(-Q^2)
& = & \hat{C}_{\lambda1}(Q)
    + \hat{C}_{\lambda2}(Q)\frac{[\hat{m}_1(Q) - \hat{m}_2(Q)]^2}{Q^2}
      \nonumber \\
&   & \qquad\qquad
    + \hat{C}_{\lambda3}(Q)\frac{[\hat{m}_1(Q) + \hat{m}_2(Q)]^2}{Q^2}
    + O\left( \frac{1}{Q^4} \right),
      \nonumber \\
\lambda^V_0(-Q^2)
& = & \hat{C}_{\lambda1}(Q)
    + \hat{C}_{\lambda2}(Q)\frac{[2\hat{m}_1(Q)^2 + 2\hat{m}_2(Q)^2]}{Q^2}
    + O\left( \frac{1}{Q^4} \right),
      \nonumber \\
\lambda^A_0(-Q^2)
& = & \hat{C}_{\lambda1}(Q)
    + \hat{C}_{\lambda3}(Q)\frac{[2\hat{m}_1(Q)^2 + 2\hat{m}_2(Q)^2]}{Q^2}
    + O\left( \frac{1}{Q^4} \right),
      \nonumber \\
\label{lambdaOPE}
\eeqa
we find
\beqa
\lefteqn{\lim_{\Lambda^2\rightarrow\infty}R_{CGN}(\Lambda^2)} \nonumber\\
& = & \lim_{\Lambda^2\rightarrow\infty}
      \frac{G_\mu}{\sqrt{2}}
      \frac{1}{2\pi i}\oint_{|s|=\Lambda^2}
      \left[ \left\{ \hat{C}_{\lambda2}(s) - \hat{C}_{\lambda3}(s)
             \right\}
             \frac{[2\hat{m}_1(s)^2 + 2\hat{m}_2(s)^2]}{-s}
           + O\left( \frac{1}{s^2}
              \right)
      \right]
\nonumber \\
& \neq & 0,
\eeqa
which shows that the DR of Ref.~\cite{Chang} will give the wrong answer for
$\Gamma(H \rightarrow \ell^+\ell^-)$, although it gives the correct answer
for $\Delta\rho$ \cite{TGW}.

Ref.~\cite{Braaten} gives the first few terms of the perturbative
expansion of the Wilson Coefficients in the running coupling
$\alpha_s(Q)$ and they are
\beqa
\hat{C}_{\lambda2}(Q)
& = & -\frac{3}{8\pi^2}
       \left[ 1 + \frac{8}{3}\frac{\alpha_s(Q)}{\pi} + \cdots
       \right],
\nonumber \\
\hat{C}_{\lambda3}(Q)
& = & -\frac{3}{8\pi^2}
       \left[ 1 + 2 \frac{\alpha_s(Q)}{\pi} + \cdots
       \right].
\eeqa
Therefore, at $O(\alpha\alpha_s)$ we find
\beqa
\lim_{\Lambda^2\rightarrow\infty}R_{CGN}(\Lambda^2)
& = & \frac{G_\mu}{\sqrt{2}}
      \frac{1}{2\pi i}\oint_{|s|=\Lambda^2} ds
      \left[ \frac{\alpha_s}{4\pi^3}
             \frac{(2m_1^2 + 2m_2^2)}{s}
      \right]
\nonumber \\
& = &  \frac{\alpha_s}{\pi}
       \left( \frac{m_1^2 + m_2^2}{4\pi^2 v^2}
       \right)
\nonumber \\
& = & \frac{\alpha_s}{\pi}
      \left( \frac{\alpha_1}{\pi} + \frac{\alpha_2}{\pi}
      \right),
\label{Changdisc}
\eeqa
where
\beq
\alpha_f \equiv \frac{m_f^2}{4\pi v^2}.
\eeq
This result coincides precisely with the
discrepancy between $\delta$
and $\delta_{CGN}(\infty)$ found in
Ref.~\cite{KScomp}.

\bigskip
\centerline{\bf V. CONCLUSIONS}
\bigskip

We have shown that in so far as QCD corrections are concerned, the
correctness of a particular dispersion relation is most easily checked by
using the Operator Product Expansion to calculate the contribution of
the circle at $|s|=\infty$. We have used the OPE to show that the
DR of Ref.~\cite{KSDR} correctly predicts the QCD
correction to the the leptonic width of the Higgs
$\Gamma(H\rightarrow \ell^+\ell^-)$ to all orders in $\alpha_s$,
whereas the DR of Ref.~\cite{Chang} introduces
an error of $(\alpha_s/\pi)\left[ (\alpha_1/\pi) + (\alpha_2/\pi) \right]$
at $O(\alpha\alpha_s)$.
\bigskip
\bigskip

\bigskip
\centerline{\bf ACKNOWLEDGMENTS}
\bigskip

This work was supported by
the United States Department of Energy under
Contract Number DE-AC02-76CH03000 and in part under
Grant Number DE-FG02-90ER40560.

\end{document}